# A DRY ELECTROPHYSIOLOGY ELECTRODE USING CNT ARRAYS

Giulio Ruffini[1], Stephen Dunne[1], Esteve Farrés [1], Josep Marco-Pallarés[1,3],
Chris Ray[1], Ernest Mendoza[2], Ravi Silva[2], Carles Grau[3]

(1) Starlab Barcelona SL, Ed. de l'Observatori Fabra, C. de l'Observatori s/n, 08035 Barcelona, Spain
(2) Nanoelectronics Center, Advanced Technology Institute, Universty of Surrey
(3) Neurodynamics Laboratory, Psychiatry and Clinical Psychobiology Department, University of Barcelona

**We describe the design of a dry electrode sensor for biopotential measurement applications (ENOBIO) designed to eliminate the noise and inconvenience associated to the use of electrolytic gel. ENOBIO uses nanotechnology to remove gel-related noise, as well as maintaining a good contact impedance to minimise interference noise. The contact surface of the electrode will be covered with an array/forest of carbon nanotubes and will also be tested with an Ag/AgCl coating to provide ionic–electronic transduction. The nanotubes are to penetrate the outer layers of the skin, the *Stratum Corneum*, improving electrical contact. We discuss requirements, skin properties, nanotube penetration and transduction, noise sources, prototype design logic and biocompatibility. A future paper will report test results.**

## INTRODUCTION

Ubiquitous monitoring of sleep quality and alertness could greatly enhance comfort, health and safety. Within the SENSATION[1] project, unobtrusive sensors, algorithms and systems are being developed for the extraction of physiological indicators associated with sleep, wakefulness and their transitions, and in particular, for monitoring our mental activity. Notably, such technologies can have a wider range of applications, including measurement of bioimpedance, brain-computer interfaces, biometrics[2], virtual/augmented reality and presence, because they open a wide window of observation to the human body and in particular to the human brain. In this paper, we discuss one of the elements the project, a nanotechnology based dry electrophysiology sensor for use in familiar environments such as the bed, car and the office—well beyond the laboratory or hospital.

Electrophysiology exploits a fundamental phenomenon of the nervous system—the biopotential. Biopotentials result from electrochemical cell activity[3] of nervous, muscular or glandular tissue (Figure 1). Impressed currents in the cellular membrane accumulate charges at the membrane, giving rise to the observed potential differences [1]. Remarkably, we can rather accurately measure biopotentials to study and monitor the function of our organism, including our heart and brain.

Since the measurement of biopotentials involves a small but finite current flow in the measuring circuit, the electrode needs to act as a transducer, transforming the ionic current into an electronic current. In non-polarizable electrodes, an electron-ion buffer coating is used, typically Ag/AgCl. In polarizable electrodes the transduction is capacitive.

Biopotential measurements are significantly noise prone, and electrodes used in demanding applications such as multi-channel EEG may require skin preparation and/or Faraday shielding in addition to electrolytic gelling. This results in long application times and a stabilization wait (required to achieve sufficient diffusion of the electrolytic gel into the skin)—and makes out-of-the-clinic applications unrealistic. Furthermore, gel will dry after a few hours and can cause skin irritation [2]. Gel and scrubbing (the mechanical removal of the skin through abrasion) are needed because of the electrical properties of skin, as we will discuss below.

---

[1] EU FP6 SENSATION Integrated Project (FP6-507231).
[2] EU FP6 HUMABIO Project (FP6-TBD)
[3] An excitable or active cell is characterized by nonlinear membrane response to depolarization, causing amplification and propagation of the depolarisation (an action potential). Excitable cells contain voltage gated ion channels.

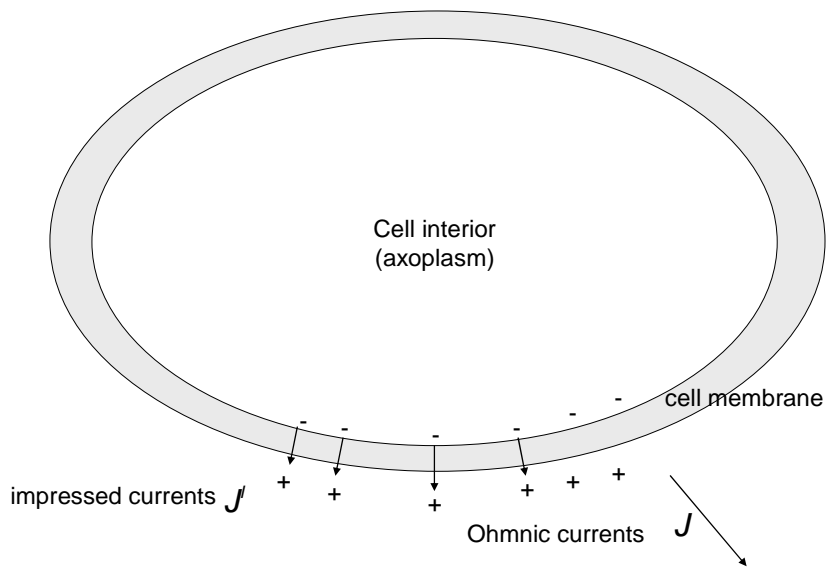

**Figure 1 Impressed currents and charges in the axon membrane**

The electrode design presented here is to eliminate the need for skin preparation and gel application. The novelty of the design is the electrode-skin interface concept, which relies on an array of large number of carbon nanotubes (CNTs) forming a penetrating brush-like structure to provide a stable electrical contact interface of low impedance. The nanostructure is designed to barely penetrate the outermost skin layer, to avoid contact with nerve cells and minimize infection risk. This approach is less aggressive than others using the same principle but working at microscales—see, e.g., [3, 4].

As a first step, we target an electrode noise level of <10μV in the frequency band 0–100Hz. Table 1 provides the SENSATION project requirements for electrophysiology measurements. The primary driver for the design of ENOBIO is EEG, with ECG and electro-oculography (EOG) as secondary drivers.

**SKIN ANATOMY AND ELECTRICAL PROPERTIES**
Because skin anatomy and associated electrical and mechanical properties crucial factors to consider in the design ENOBIO, we begin by discussing them in some detail. The skin has a layered architecture which can vary greatly depending on its body location. In general, skin consists of three layers: epidermis, dermis and hypodermis. The outer layer of skin is the epidermis, with a thickness varying from 50 μm in the scalp to 100 μm in the sole. The *Stratum Corneum* (SC), the outer layer of the epidermis, has a thickness of 10 to 20 μm. As other bodily tissues, skin is a resistive medium [1].

|  | NOISE | AMPLITUDE | SAMPLING | QUANTIZATION |
|---|---|---|---|---|
| **EEG** | < 1μVpp or < 0.1μV$_{eff}$ within 0.1-20 Hz | ± 1000μV | 0.01 Hz – 90 Hz with a sampling frequency ≥ 300 Hz (Opt. 500 Hz) | 16 bit A/D |
| **EOG** | < 2μVpp or < 0.2μVeff within 0.1-20 Hz | ± 2000μV | 0.01 Hz – 90 Hz with a sampling frequency ≥ 300 Hz (Opt. 500 Hz) | 16 bit A/D |
| **ECG** | < 5μVpp or < 0.5μVeff within 0.1-20 Hz | ± 2000μV | 0.1-100 Hz with a sampling rate ≥ 300 (opt. 500 Hz) | 16 bit A/D |
| **EMG** | < 1μVpp or < 0.1μVeff within 0.1-20 Hz | ± 1000μV | 15-2000Hz. | 16 bit A/D |

**Table 1 Summary of optimal requirements for electrophysiology measurements**

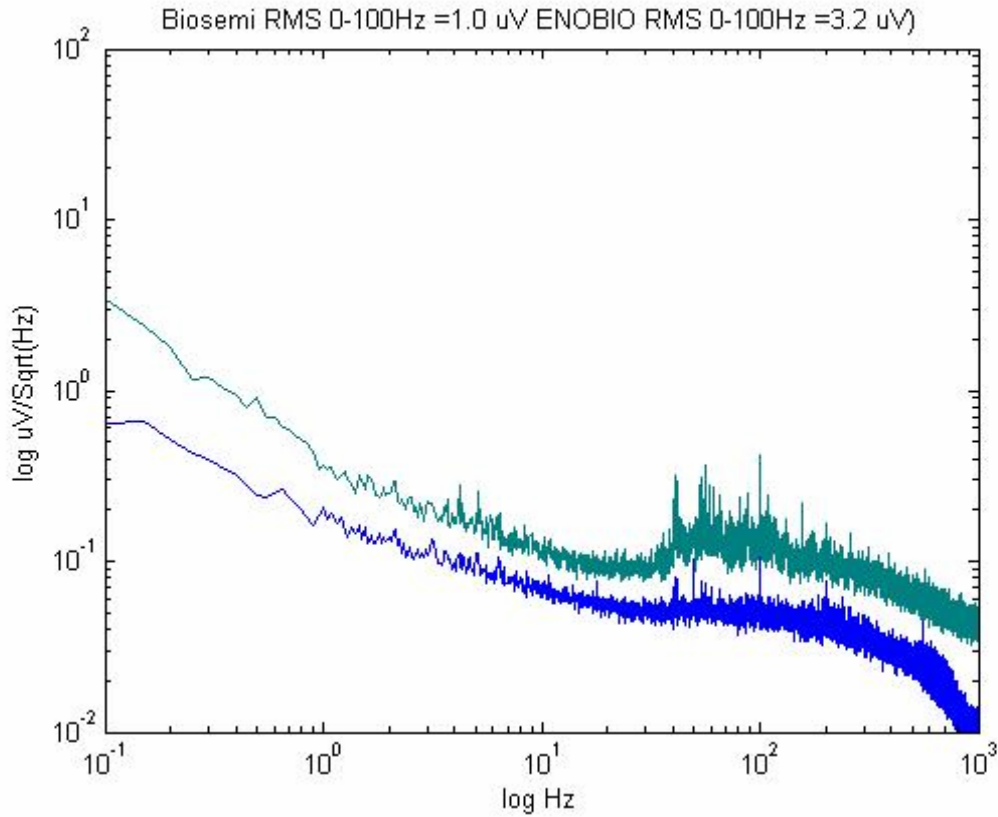

**Figure 2** Noise spectra for initial tests with a state-of-the-art electrode (Biosemi[4], in blue) compared with an ENOBIO prototype without a coating (see [4]). Horizontal axis is the log10 of frequency (Hz), and the vertical log10 μV/sqrt(Hz). For the Biosemi electrode, the voltage RMS is ~1 μV after filtering from 0.1 to 100 Hz while for the ENOBIO electrode RMS is ~3 μV. When measuring electrode noise Biosemi was referenced to a Biosemi electrode and ENOBIO to an ENOBIO electrode.

The SC is a very important fluid barrier consisting mostly of constantly renewing dead skin cells. Unlike the strata underneath, which bathe in electrolytic interstitial fluid, it is therefore highly resistive. It is because of the highly resistive character of the SC that electrolytic gels and skin scrubbing are needed in electrophysiology: in order to obtain a good contact between the electrode and the electrolytic tissue below, the SC must be bypassed, removed or its electrical properties altered.

**NOISE SOURCES**
There are various sources of noise associated to electrophysiology measurements (see Figure 2). Some of these sources can be classified as "contact" noise and are discussed in [5], while others are more properly called "environment" noise [6].

**Contact noise**
This is the noise originating at unstable interfaces. Examples include the electrode-gel interface and gel-SC interfaces. The SC-gel interface is the most important source of noise in a wet electrode. In penetrating type of electrodes, such as the one discussed here, we achieve a direct electrode-skin interface.

Half-cell (HC) noise: This type of noise is associated to the dynamics of the charge density function. The origin of the charge distribution in an overall neutral object is polarization. Since material electron affinity varies, polarization occurs at interfaces and gives rise to a small electric field and potential step. Polarization dynamics translate into voltage noise, which is produced at several interfaces, including the relatively stable electrode-gel and the more unstable gel-skin (if gel is present), or electrode-epidermis (as in ENOBIO)—which should be more stable. Half-cell noise in traditional electrodes is probably mostly due to the diffusion of gel in the SC—a dynamic process in a very inhomogeneous material. HC noise can also induce motion artifacts (see below).

---
[4] Biosemi Active 2 system, http://www.biosemi.com

Overpotential noise: Charges accumulate when currents flow across the electrode interface. This effect, minimized with the use of non-polarizable electrodes are and/or if high input impedance amplification, gives rise to motion artifacts which are of particular importance to EEG (they occur at about the same low frequencies of biological interest of 0.1 to 100 Hz). Motion artifacts arise from mechanical disturbances of the charge distribution at the electrode-electrolyte interfaces. In traditional electrodes, the skin-gel interface is again mostly at fault.

Thermal noise: Thermal noise is proportional to impedance and bandwidth [5]: $<V_{th}^2> = 4kTBR$, where k is the Boltzman constant, T is temperature in K, B is Bandwidth in Hz and R is the electrode Impedance in Ohm. With our target impedance of 20 k$\Omega$, at room temperature (T=300 K) and 100 Hz bandwidth, this amounts to ~0.2 µV.

**Environment noise**

Amplifier noise: This is noise intrinsic to the amplification process, and is described by the amplifier noise figure. This type of noise is typically rather small in comparison to the other components. As discussed in [3], neither thermal nor amplifier noise play a significant role in electrode noise.

Interference noise: This is noise due to the presence of external EM fields coupling to the subject or measuring apparatus. In the following, we refer to Figure 3.

Body-EM coupling noise is due to electric-field capacitive coupling between the power lines and subject. The result is a common-mode voltage on the body that induces currents through the electrodes. These voltages are of the order of 10-20 mV. For two locations A and B we would have (for a large input impedance Z), $V_A - V_B = V_{CM} (Z_A - Z_B)/Z$, where $Z_A$ and $Z_B$ are the electrode-skin contact impedance at two measurement points A and B. Therefore if Z is very large the noise voltage $V_A-V_B$ is minimized. We can express this relationship by $\Delta V = \Delta Z\, V_{CM}/Z$. For example, a typical input impedance is roughly Z=300 M$\Omega$. Taking $V_{CM}$=20 mV and requiring the error to remain below 1 µV, we need $\Delta Z$ < 30 k$\Omega$.

Lead-EM coupling: exposed leads can also couple to the power lines and other EM sources. For example a current of the order of a few nA is produced by cables several meters in length (6 nA are quoted in [6] for a 9 m lead). This will cause voltage differences if the impedances of the different electrodes are not equal, $V_A-V_B = I (Z_A-Z_B)$, or $\Delta V = I\, \Delta Z$. For a current of 6 nA, with impedance deltas of the order of 20 k$\Omega$, we are led to $\Delta V$= 120 µV.

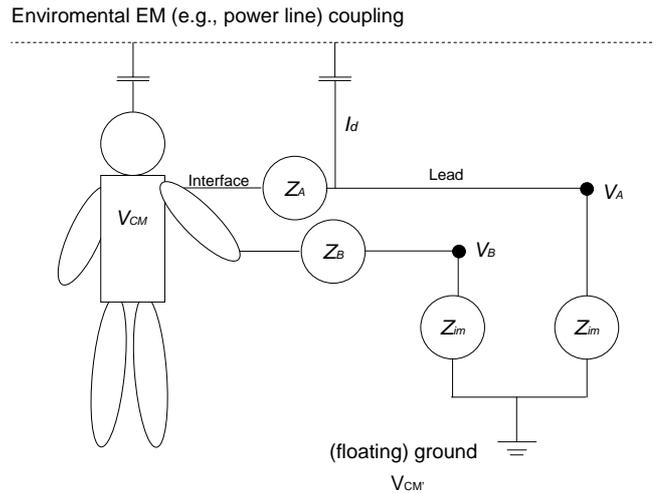

**Figure 3 Interference noise due to coupling to environmental EM fields. For high enough input impedance as compared to the contact impedance, the voltages at A and B are basically those at the skin.**

For these reasons, we shall implement on an "active electrode" approach, with on-site amplification [6]: the "lead cables" are extremely short with Active Electrodes, which minimizes the impact of this noise source.

Finally, we wish to emphasize that, although traditionally reducing contact impedance has been as an important element for good signal quality, with the advent of high input impedance, on-site (local) amplification, it has become significantly less important. As long as contact impedance remains below a reasonable value, e.g., ~100 kΩ, interference noise will not be large. With modern non-polarizable electrodes the main source of noise are variable half-cell potentials, especially arising at the liquid-liquid junction (gel-skin). We expect to greatly reduce it with a penetrating active electrode concept.

**ELECTRODE CONCEPT**
Despite uncertainty in skin properties at nanoscales, our initial sensor paradigm assumes stiff CNTs arrays partly penetrating the SC.

Since the electrode is to transduce ionic currents into electronic currents at low frequencies, the initial design will be non-polarizable, with an AgCl coating. Ag/AgCl is biocompatible and acts like an excellent transducer, allowing currents to pass freely without overpotentials (reversible REDOX), thus reducing motion artefacts. A polarizable electrode with no coating shall also be tested, however—CNTs are naturally inert and modern electronics can provide very high input impedance.

We seek an overall contact impedance of 20 kΩ, although 100 kΩ will be considered sufficient. This means that the CNT interface alone should have a low impedance, and in particular that the CNT, CNT-substrate contact, and CNT-growth substrate resistivities should all be low.

The CNT interface will be initially be mounted on commercial active electrode and connected to electrophysiology recording equipment, where it will be tested and compared to existing "wet" sensors. The equipment we use is the Biosemi Active 2 [8]—a state of the art, research oriented and flexible platform.

**CNT INTERFACE ARCHITECTURE**
To develop the electrophysiological sensor, multiwalled carbon nanotubes (MWCNTs) have been selected, because they meet our good conductivity requirements and are stiffer than single-walled carbon nanotubes. The CNT array architecture has been designed to achieve penetration of only the outer parts of the SC, and to avoid a "bed of nails/fakir" effect. The catalyst used to grow the first set of CNT arrays is Ni, although more biocompatible alternatives are now being used.

**Mechanical aspects**

As the electrode concept described here employs a skin penetration paradigm, we need to analyze the resistance of the SC to penetration by needle-like structures. We address here the potential buckling of CNTs on contact with the skin, an effect that would prevent penetration. To understand if the tubes will puncture the skin or buckle, we use Euler's formula for the critical pressure a tube can withstand [9]. This models the tube as a hollow cylinder of external radius $R_1$, internal $R_2$, length L and Young Modulus, which we take here to be 1 TPa [10]:

$$P_{CR} = \pi^2 \frac{EI}{L^2}, \text{ with } I = \pi(R_1^4 - R_2^4)/64.$$

Using these, we see that the buckling pressure of our design will be of the order of 50 MPa (the inner radius is assumed half of outer).

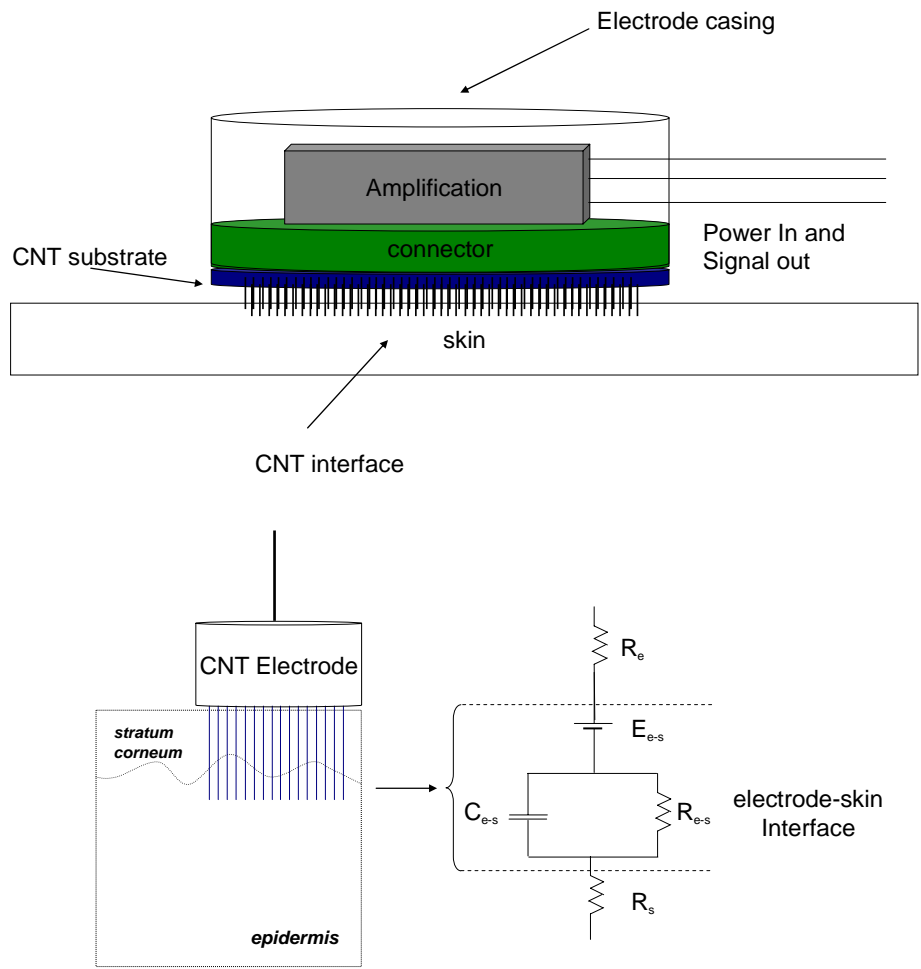

**Figure 4 Electrode interface concept (top) and equivalent circuit (bottom), showing the inherent electrode resistance, the contact related half-cell potential and parallel capacitance-resistance, and epidermis resistance.**

As we now discuss, a macroscopic skin puncture model predicts a very large resistance to penetration at small scales. Following the terminology in [9], we have $J_{IC}$=2500 J m$^{-2}$ (this is the *toughness parameter* of skin, opposing mode I cracking), α=9.0 (skin strain hardening exponent) and μ=0.11 MPa (skin infinitesimal shear modulus). In order to estimate the critical pressure to puncture skin based on [9], we compute the scale-sensitive toughness coefficient, $J_{IC}/\mu R$. The critical pressure increases rapidly with the toughness coefficient.

The skin toughness coefficients associated to the size of a MWCNT (R=2x10$^{-7}$m) and a standard 12.7 mm long hypodermic needle (R=7x10$^{-4}$m) are, respectively, ~114,000 and ~30: at nanoscales the skin is extremely tough (driven by mode I linear cracking). This in turn leads to an estimate of about 1 GPa critical pressure. However, we question the model's applicability to nanoscales and with real skin (which cannot be assumed homogeneous at nanoscale). If cracking does not occur (a 1d process consuming a lot of energy at small scales), then the smaller μ is the reference critical pressure. Nanoscale skin features may present softer spots than those observed, as average quantities, at macroscale. Nonetheless, CNTs should be engineered to be stiff as possible, considering the requirement of a minimal length of 10 μm.

**Electrical aspects**
ENOBIO specify 20 kΩ as an acceptable contact impedance. A reasonable value for the resistance of a multi-walled carbon nanotube is ~100 kΩ – 2000 kΩ [11]. We assume here a CNT-skin contact impedance of $R_{CNT}$ = 1 MΩ. Since the total resistance of *N* nanotubes contacting the skin in parallel is $R_T = R_{CNT}/N$, it follows that to obtain the specified contact impedance ~50 nanotubes should penetrate. To increase the safety margin of the design let us assume that only 10 % of the tubes contact the skin. Then, the number of CNTs needed in the array is of the order of ~500. As the design of the sensor specifies an active area of 2 x 2 mm$^2$, the minimal density of tubes needed is ~125 CNT/ mm$^2$. Therefore the spacing between tubes should be less than 12 CNT / mm, or ~1 CNT/80 μm. From the mechanical point of view, this spacing is also satisfactory to avoid a or bed-of-nails effect.

Finally,we have also considered the potential redox saturation of the Ag/AgCl coating by the sustained currents. A back-of-the-envelope calculation shows that given the dimensions of the CNTs, the number of them expected to make contact, and the current amplitudes, there does not appear to be a danger of saturation.

**CNT INTERFACE GROWTH**
To grow the CNTs we used the vapor based catalytic method also known as the Chemical Vapor Deposition (CVD), that is based on the breaking down of carbon containing gases, e.g., acetylene, ethylene, carbon monoxide, etc., over a catalyst material which is typically a transition metal. The breaking down of the hydrocarbon gases are usually achieved by thermal means or assisting plasma. Nanotubes are grown over the catalyst and collected when the system cools down back to room-temperature. Due to the lower energy involved in the CVD method as compared to the arc discharge or laser ablation techniques, CVD grown nanotubes are usually more defective and therefore appear curly if no additional steps are taken to align it during/before the growth. It has been known since the early '70s that carbon filaments can be synthesized by catalytically breaking down hydrocarbon gases over hot surfaces [12]. The filaments observed in the past have diameters from 10nm to 0.5μm and are tens to hundreds of microns long.

According to this mechanism, the growth of the filaments starts off with the decomposition of the hydrocarbon at the 'front' surface of the metal catalyst particle, producing hydrogen and carbon, which then dissolves in the metal. The dissolved carbon then diffuses through the particle, to be deposited on the trailing "cool" face. The precipitation of carbon from the saturated metal particle leads to the formation of the filament. The filament growth then proceeds until the carbon feedstock is cut off, the catalyst particle is covered or thermal energy is removed. Using controlled atmosphere electron microscopy (CAEM), Baker and co-worker were able to determine the kinetics of the process, showing that the activation energy for the filament growth was about the same as the activation energy for bulk carbon diffusion in nickel [12]. According to them, the main driving force is the temperature and the concentration gradient with the rate limiting step being the diffusion of carbon through the catalyst particle.

However, carbon nanotube growth with room temperature held substrates has been recently demonstrated [13, 14]. The energy is supplied by a plasma produced by either a RF or microwave source. The substrate is cooled to ensure that it does not reach a macroscopic temperature above 60° C. In these studies we have been able to couple the plasma energy efficiently to the Ni particles in order to substitute the plasma energy instead of the conventional high temperature substrates. It is true that the Ni-C eutectic temperatures must be reached at the localized surfaces in contact with the plasma sheath, but the bottom end of the Ni "tear-drop" lifted away from the substrate rapidly quenches the fast diffusing surface carbon and precipitate in the form of CNT.

The size of the growth catalyst is of the order of ~ 50–100 nm–to obtain a single CNT from every Ni dot. The pattern of such size Ni arrays cannot be performed using the standard light lithographic techniques because the resolution of these is limited to dots of the order of 800 nm, but electron beam lithography can be used. The resolution of this technique is much higher than the UV lithography reaching dots of sizes around 20 – 30 nm. The technique is more accurate because uses an electron beam to expose a special resist. Alternatively, we will also test a non-homogeneous CNT distribution with the equivalent average spacing (uniformity is not needed).

The Ag/AgCl coating technique that has been chosen for the first investigation is laser ablation. A target made of melted silver chloride was ablated with a 248 nm laser (6 J/cm$^2$). The ablated AgCl vaporized and decomposed into metallic silver and gaseous chlorine and the vapours of metal were deposited on the outside wall of CNTs used as a template. Different numbers of laser shots were used to asses the influence on the deposited size of silver crystals as well as the coverage of the nanotube surface. Figure 6(a) shows a Scanning Electron Microscope (SEM) image of a CNT array covered with Ag. Figure 6 (b) presents a magnification of the tubes that shows the Ag particles coating the walls extremely uniformly on the CNTs.

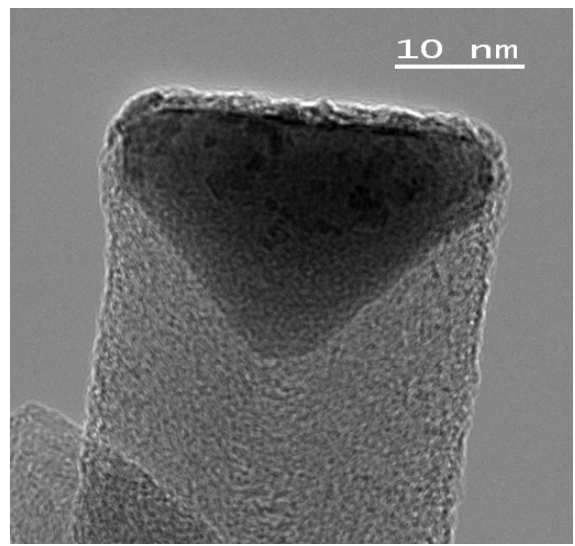

**Figure 5 Transmission Electron Microscopy micrography of a room temperature synthesised carbon nanotube. The image shows the Ni catalyst particle at its tip**

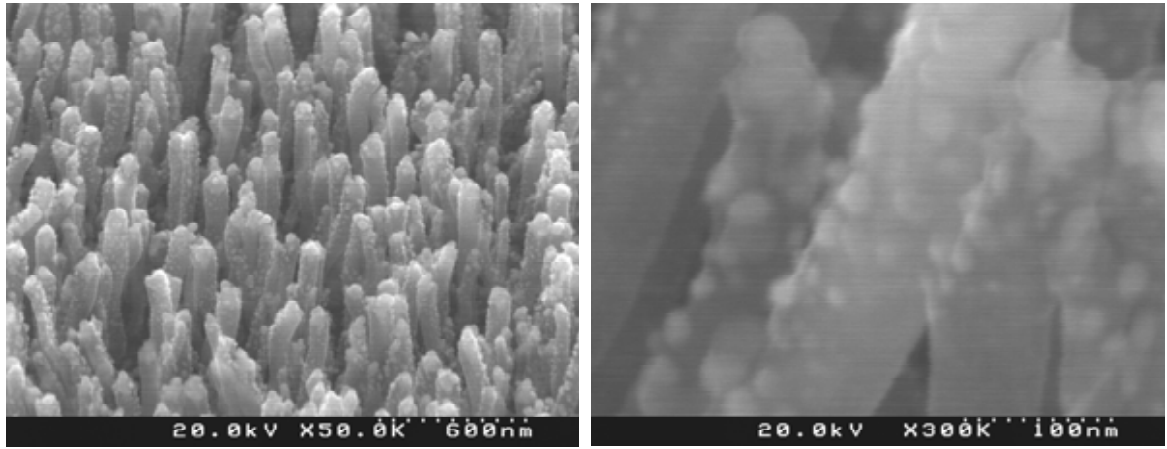

**Figure 6 (a) SEM image showing an array of CNTs coated with Ag. (b) Magnification showing small Ag droplets covering the walls of the tubes.**

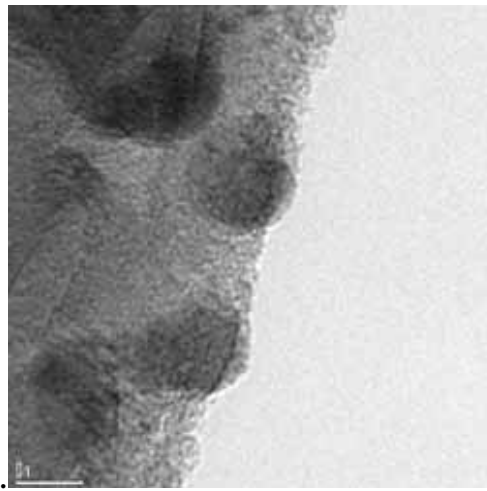

**Figure 7 Ag droplets on the CNTs**

Figure 6 (b) shows small droplets covering the walls of the CNTs. Currently; the number of laser shots is being increased to achieve a homogeneous coverage of all the surface of the CNT. Figure 7 shows a Transmission Electron Microscopy (TEM) image with a region subject to high magnification of the wall of a CNT. The metallic particles have well pronounced crystal structure with the lattice distance typical for Face Centred Cell (FCC) of silver. The size of the particles appears to be dependent of the number of laser shots—an effect currently under investigation. After ablation, the surface is chlorinated to produce AgCl layer on top of the Ag.

The substrate used for the first prototype is highly doped Si and therefore highly conducting. This substrate is currently a standard substrate for the growth of CNTs and the best option for the first prototype of our device. Concerning the catalyst, during this first period we have setup a new DC-RF sputtering system for the growth of thin films. The advantage of using such a system in comparison with metal evaporators is that the adhesion of the catalyst film is higher and therefore the adhesion of the CNTs to the substrate will be enhanced.

Once the concept prototype has been constructed and thoroughly tested we can aim to miniaturize it to reduce the overall size of the electrode. The possibility of growing CNTs on flexible substrates will also be investigated.

**SAFETY ISSUES**
ENOBIO is designed to penetrate only the skin outer layer with very small diameter punctures. This should lower infection risk significantly in comparison to micro-needle approaches. However, despite the very small diameter and length of CNTs, infection is still a possibility: small viruses and prions could penetrate beyond the SC. However, infection risk is not an unusual threat in electrophysiology practice and could be dealt with by the use of disposable sensors.

With regards to biocompatibility, we recall that the three main contact sites of the body with the environment are skin, lungs and intestinal tract. Inhalation is a direct way to introduce a strange particle into the body. Although it is well known that atmospheric aerosols contain a great variety of particles, in a wide range of sizes, there is a great interest in knowing the effect of the smallest ones. It has been shown [15] that these ultra-fine particles could be the most dangerous ones, crossing the blood-brain barrier and impacting the Central Nervous System. Because of their characteristics, CNTs share properties with asbestos, a material that has demonstrated a great toxicity on lungs. Two independent studies [16, 17] reported the finding of granulomas and some interstitial inflammation. The first research group worked with massive inhalation of CNTs in a group of rats, concluding that the findings of multifocal granulomas may not have physiological relevance. The second group instilled a suspension of nanotubes directly into the lungs of mice and argued that if CNTs reach the lungs, they are more toxic than carbon black and quartz. Finally, another group reports that CNTs do not exhibit effects similar to asbestos [18].

Several studies have been carried out to assess skin-CNT interaction. The first and most directly related to the present concept [19] performed tests on rabbits, and failed to find any signs of health hazards related to skin irritation and allergic risks. On the other hand, [20] found that exposure to SWCNT resulted in ultrastructural and morphological changes in cultured skin cells and concluded that dermal exposure to unrefined SWCNT may lead to dermal toxicity due to accelerated oxidative stress in the skin. This study was performed during 18 h of exposure of SWCNT to cultured human epidermal keratinocytes. Note that exposure of SWCNTs to cultured skin cells is much more invasive than natural skin contact, and this may explain the different results. And it should also be noted that the presence of iron in these SWCNTs (from the catalyst) may have been the cause of accelerated oxidative stress. Our design includes a biocompatible coating, which should limit potential harmful effects.

Clearly, any CNT based technology should ensure minimal dispersion of CNTs to the environment, and future studies are needed to assess the risk of CNTs. For this reason we are studying mechanisms to limit the possibility of CNT-substrate detachment and planning on performing a very limited set of human tests to validate the concept. These will be discussed in a future publication.

**CONCLUSIONS**
The design of a CNT-based electrophysiology electrode is a fascinating and challenging multi-disciplinary exercise involving analysis of requirements, skin and CNT mechanical and electrical properties at nanoscale, electrochemistry and biosafety issues. As such it can best be carried out within the scope of multi-disciplinary integrated projects bringing together the expertise of several key partners.

We have analyzed here the requirements and proposed a new concept of dry electrode based on the use of CNTs. The problems for practical applications of electrophysiology arise from the need for gel application, scrubbing and/or Faraday caging to minimize noise. The source of the most difficult-to-handle noise sources is the SC-gel interface, and we have proposed a way to bypass it in a painless way to access the more conductive skin layers below. The proposed electrode is to penetrate the SC at nanoscales in order to reduce measurement noise painlessly, enabling quality recording of biopotential signals such as EEG without gel or skin scrubbing.

Testing of the electrodes can reveal, through observed effects on contact impedance and reduction in noise, if penetration is taking place. If skin preparation (scrubbing) has little effect on signal quality or observed electrodes-skin impedance we can infer that the CNTs are indeed penetrating the SC. In a future publication we will report on testing plans, results and design iterations, with the contributions from other partners in the project.

**ACKNOWLEDGEMENTS**
ENOBIO is developed under the European Integrated Project SENSATION (FP6-507231), and partly supported by the Starlab's Kolmogorov project. Chris Ray, on sabbatical leave from St. Mary´s College (CA, USA), was at Starlab as an invited visiting scholar. G. Ruffini wishes to thank A. Kasumov and W. de Brower for initial discussions on ENOBIO as part of Starlab's SleepLab initiative, N. Fleck for useful discussions on the applicability his macroscopic penetration models to nanophysics, and C. Meeting for valuable comments on sources of electrode noise. All Starlab authors have contributed significantly and the Starlab author list has been ordered randomly.

**Corresponding Author**: Giulio Ruffini . Edifici de l'Observatori s/n. 08035. Barcelona. Tel +34-932540966 , fax +34-932126445. e-mail: giulio.ruffini@starlab.es